\documentclass[12pt,a4paper]{amsart}
\usepackage[hypertex]{hyperref}
\usepackage{subfigure,caption2}
\usepackage{amssymb}
\usepackage{color}
\usepackage{graphicx}
\usepackage[rflt]{floatflt}
\numberwithin{equation}{section}
\newcommand{\smallpagebreak}{{\par\vspace{2 mm}\noindent}}

\newcommand{\dsize}{\textstyle}

\newcommand{\R}{{\mathbb R}}

\newcommand{\Z}{{\mathbb Z}}
\newcommand{\N}{{\mathbb N}}
\newcommand{\C}{{\mathbb C}}

\newcommand{\re}{{\rm Re}\,}
\newcommand{\im}{{\rm Im}\,}

\newcommand{\res}{{\rm res}\, }
\newcommand{\Const}{{\rm Const}}
\theoremstyle{plain}
\newtheorem{Th}{Theorem}[section]
\newtheorem{Le}{Lemma}[section]

\theoremstyle{definition}

\newtheorem{Def}{Definition}[section]
\title{The width of resonances for slowly varying perturbations of
  one-dimensional periodic {Schr{\"o}dinger} operators}
\author{Fr{\'e}d{\'e}ric Klopp} \author{Magali Marx}
\address[Fr{\'e}d{\'e}ric Klopp]{LAGA, Institut Galil{\'e}e, U.R.A 7539 C.N.R.S,
  Universit{\'e} Paris-Nord, Avenue J.-B.  Cl{\'e}ment, F-93430
  Villetaneuse, France\\ et\\ Institut Universitaire de France}
\email{\href{mailto:klopp@math.univ-paris13.fr}{klopp@math.univ-paris13.fr}}
\address[Magali Marx]{Institut Fourier, 100 rue des Maths, BP 74, 38402 Saint-Martin d'Heres cedex, France}
\email{\href{mailto:murat@math.univ-paris13.fr}{murat@math.univ-paris13.fr}}
\begin{document}
\begin{abstract}
  In this talk, we report on results about the width of the resonances
  for a slowly varying perturbation of a periodic operator. The study
  takes place in dimension one. The perturbation is assumed to be
  analytic and local in the sense that it tends to a constant at
  $+\infty$ and at $-\infty$; these constants may differ. Modulo an
  assumption on the relative position of the range of the local
  perturbation with respect to the spectrum of the background periodic
  operator, we show that the width of the resonances is essentially
  given by a tunneling effect in a suitable phase space.
  \vskip.5cm
  \par\noindent   \textsc{R{\'e}sum{\'e}.}
  Dans cet expos{\'e}, nous d{\'e}crirons le calcul de la largeur des
  r{\'e}sonances de perturbations lentes d'op{\'e}rateurs de Schr{\"o}dinger
  p{\'e}riodiques. Cette {\'e}tude est uni-dimensionnelle. Les perturbations
  lentes consid{\'e}r{\'e}es sont analytiques et locales au sens o{\`u} elles
  tendent vers une constante en $+\infty$ et en $-\infty$ ; ces deux
  constantes peuvent toutefois {\^e}tre diff{\'e}rentes. Sous des hypoth{\`e}ses
  ad{\'e}quates sur la position relative de l'image de la perturbation
  locale par rapport au spectre de l'op{\'e}rateur de Schr{\"o}dinger
  p{\'e}riodique, nous d{\'e}montrons que la largeur des r{\'e}sonances est donn{\'e}e
  par un effet tunnel dans un espace de phase ad{\'e}quat.
\end{abstract}
\setcounter{section}{-1}
\maketitle
\section{Introduction}
\label{sec:introduction}
The present talk is devoted to the analysis of the family of
one-dimensional quasi-periodic Schr{\"o}dinger operators acting on
$L^2(\R)$ defined by
\begin{equation}
  \label{family}
  H_{\zeta,\varepsilon}=-\frac{d^2}{dx^2}+V(x)+W(\varepsilon x+\zeta).
\end{equation}
We assume that
\begin{description}
\item[(H1)] $V:\ \R\to\R$ is a non constant, locally square
  integrable, $1$-periodic function;
\item[(H2)] $\varepsilon$ is a small positive number;
\item[(H3)] $\zeta$ is a real parameter;
\item[(H4)] $W$ is a potential that is real analytic in a conic
  neighboring the real axis that admits a limit at $+\infty$ and at
  $-\infty$; the precise assumption is stated in
  section~\ref{sec:assumptions}.
\end{description}
\noindent As $\varepsilon$ is small, the operator~\eqref{family} is a
slow perturbation of the periodic Schr{\"o}dinger operator
\begin{equation}
  \label{Ho}
  H_0=-\frac{d^2}{dx^2}+V\,(x)
\end{equation}
acting on $L^2(\R)$. \\
When $V\equiv0$, the operator $H_{\zeta,\varepsilon}$ is independent
of $\zeta$ (up to a unitary equivalence, a translation actually) and
becomes a semi-classical Schr{\"o}dinger operator (by a simple rescaling
of the variable $x$). The resonances for such operators have been the
subject of a vast literature over the last 20 years; a detailed review
and many references can be found in the lecture notes~\cite{Sjo:06}
and in the papers~\cite{MR1957536,MR2000d:58051}.\\
The problem of computing the resonances for slowly varying
perturbations of a periodic Schr{\"o}dinger operator has been much
less studied. The papers~\cite{MR2003h:35189,MR1962356} deal with
the multi-dimensional case; asymptotic formulas are obtained for
the real part of the resonances. This is done using trace formulas
and the imaginary parts of the resonances have not been computed.
The papers~\cite{MR99d:34164,MR2002j:35018} are devoted to the
study of Stark-Wannier resonances in a small electric field.
\par It is well known that the eigenvalues and resonances for
$H_{\zeta,\varepsilon}$ near an energy $E$ depends very much on the
region of energy one is studying. Les us assume $W$ tends to $0$ at
both ends of the real axis. Then, the absolutely continuous spectrum
of $H_{\zeta,\varepsilon}$ is the spectrum of $H_0$. In this case, we
compute the resonances of $H_{\zeta,\varepsilon}$ near energies inside
the spectrum of $H_0$. Actually, for $W$, we also consider the case
when the limits of $W$ at $+\infty$ and $-\infty$ differ. We obtain
that the real parts of the resonances are given by a quantization
condition interpreted naturally in the adiabatic phase space (see
section~\ref{sec:heuristics}). The imaginary parts of the resonances
are then given by an exponentially small tunneling coefficient.
\par Though the results depend crucially on the case dealt with, from
the technical point of view, both cases can be dealt with in a similar
fashion. To study~\eqref{family}, we use the asymptotic method
developed in~\cite{MR2002h:81069,MR2097997,Mar:04} for the analysis of
slow perturbations of one-dimensional periodic equations.\\

\section{The results}
\label{sec:results}
We first recall some of elements of the spectral theory of
one-dimensional periodic Schr{\"o}dinger operator $H_0$; then, we present
our results.
\subsection{The periodic operator}
\label{sec:periodic-operator}
For more details and proofs we refer to section~\ref{S3} and
to~\cite{Eas:73,MR2002f:81151}.
\subsubsection{The spectrum of $H_0$}
\label{sec:son-spectre}
The spectrum of the operator $H_0$ defined in~\eqref{Ho} is a union of
countably many intervals of the real axis, say $[E_{2n+1},\,E_{2n+2}]$
for $n\in\N$ , such that
\begin{gather*}
  E_1<E_2\le E_3<E_4\dots E_{2n}\le E_{2n+1}<E_{2n+2}\le \dots\,,\\
  E_n\to+\infty,\quad n\to+\infty.
\end{gather*}
This spectrum is purely absolutely continuous. The points
$(E_{j})_{j\in\N}$ are the eigenvalues of the self-adjoint operator
obtained by considering the differential polynomial~\eqref{Ho} acting
in $L^2([0,2])$ with periodic boundary conditions (see~\cite{Eas:73}).
The intervals $[E_{2n+1},\,E_{2n+2}]$, $n\in\N$, are the {\it spectral
  bands}, and the intervals $(E_{2n},\,E_{2n+1})$, $n\in\N^*$, the
{\it spectral gaps}. When $E_{2n}<E_{2n+1}$, one says that the $n$-th
gap is {\it open}; when $[E_{2n-1},E_{2n}]$ is separated from the rest
of the spectrum by open gaps, the $n$-th band is said to be {\it
  isolated}.
\smallpagebreak From now on, to simplify the exposition, we suppose
that
\begin{description}
\item[(O)] all the gaps of the spectrum of $H_0$ are open.
\end{description}
\subsubsection{The Bloch quasi-momentum}
\label{sec:le-quasi-moment}
Let $x\mapsto\psi(x,E)$ be a non trivial solution to the periodic
Schr{\"o}din\-ger equation $H_0\psi=E\psi$ such that, for some $\mu\in\C$
and all $x\in\R$, $\psi\,(x+1,E)=\mu \,\psi\,(x,E)$. This solution is
called a {\it Bloch solution} to the equation, and $\mu$ is the {\it
  Floquet multiplier} associated to $\psi$. One may write
$\mu=\exp(ik)$; then, $k$ is the {\it Bloch quasi-momentum} of the
Bloch solution $\psi$.
\smallpagebreak The mapping $E\mapsto k(E)$ is an analytic
multi-valued function; its branch points are the points $E_1$, $E_2$,
$E_3$, $\dots$, $E_n$, $\dots$. They are all of ``square root'' type.
\smallpagebreak The dispersion relation $k\mapsto{\mathbf E}(k)$ is
the inverse of the Bloch quasi-momentum.  We refer to section~\ref{S3}
for more details on $k$.
\subsection{The assumptions on $W$ and the analytic continuation of
  the resolvent}
\label{sec:assumptions}
We now make assumption (H4) more precise. We introduce the following
notation: for $C_0>0$, define $\mathcal{C}_{C_0}$ to be the cone
\begin{equation*}
  \mathcal{C}_{C_0}=\{z\in\C;\ C_0|\text{Im}\,z|\leq
  1+|\text{Re}\,z|\}.
\end{equation*}
We assume
\begin{description}
\item[(H4a)] there exists $C_0>0$ such that $W:\
  \mathcal{C}_{C_0}\to\C$ is real analytic, non constant;
\item[(H4b)] there exist $(W_+,W_-)\in\R^2$ and $s>1$ such that, for
  any $C_1>C_0$
  \begin{equation}
    \label{eq:4}
    \sup_{\substack{z\in\mathcal{C}_{C_1}\\\re z>0}}
    \left[|z|^s|W(z)-W_+|\right]+
    \sup_{\substack{z\in\mathcal{C}_{C_1}\\\re z<0}}
    \left[|z|^s|W(z)-W_-|\right]<+\infty
  \end{equation}
\end{description}
So $W$ is short range in cones neighboring $+\infty$ and $-\infty$. It
is well known that, when $H_0$ is the free Laplace operator, this
assumption guarantees that the resolvent of $H_0+W$ can be analytically
continued from the upper half-plane through the spectrum of $H_0$ as a
mapping from $L^2_{r_0}(\R)$ to $L^2_{-r_0}(\R)$ (for some $r_0>0$)
(see~\cite{MR871788,MR1957536} and references therein). Here, for
$r\in\R$, we have defined
\begin{equation*}
  L^2_r=\{u\in L^2_{\text{loc}}(\R);\ e^{r|\cdot|}u(\cdot)\in
  L^2(\R)\}.
\end{equation*}
For $\text{Im}\,E>0$ and $\zeta\in\R$, the resolvent of
$H_{\zeta,\varepsilon}$ at energy $E$, that is $E\mapsto
R(E,\zeta,\varepsilon):= (H_{\zeta,\varepsilon}-E)^{-1}$ is a function
valued in the bounded operators on $L^2(\R)$; moreover, under
assumption (H4), a simple resolvent expansion shows that, for any
$(E_0,\zeta_0)\in\{\text{Im}\,E>0\}\times\R$, it is analytic in a
neighborhood of $(E_0,\zeta_0)$. We prove
\begin{Th}
  \label{thr:2}
  Assume (H1) -- (H4) are satisfied. Pick $E_0\in\R$ such that either
  $E_0-W_+\in\sigma(H_0)\setminus\partial\sigma(H_0)$, or
  $E_0-W_-\in\sigma(H_0)\setminus\partial\sigma(H_0)$, or
  both hold.\\
  Then, there exist $\rho_0>0$, $r_0>0$, $\varepsilon_0>0$ and a
  complex valued function $\Delta:\ (E,\zeta,\varepsilon)\mapsto
  \Delta(E,\zeta,\varepsilon)$ defined on $D(E_0,r_0)\times
  \{\zeta;|\text{Im}\,\zeta|<r_0\}\times(0,\varepsilon_0)$ such that,
  for $\varepsilon\in(0,\varepsilon_0)$,
  \begin{itemize}
  \item the mapping $(E,\zeta)\mapsto\Delta(E,\zeta,\varepsilon)$ does
    not vanish on $\{E;\ |E-E_0|<r_0,\ \text{Im}\,E>0\}\times\R$;
  \item the mapping $(E,\zeta)\mapsto\Delta(E,\zeta,\varepsilon)$ is
    analytic on $D(E_0,r_0)\times\{\zeta;|\text{Im}\,\zeta|<r_0\}$;
  \item the mapping $(E,\zeta)\mapsto \Delta(E,\zeta,\varepsilon)
    R(E,\zeta,\varepsilon)$ can be continued analytically from $\{E;\
    |E-E_0|<r_0,\ \text{Im}\,E>0\}\times\R$ to $\{E;\ |E-E_0|<r_0\}
    \times\{\zeta;|\text{Im}\,\zeta|<r_0\}$ as a non-vanishing bounded
    operator from $L^2_{\rho_0}(\R)$ to $L^2_{-\rho_0}(\R)$;
  \item for any $E\in D(0,r_0)$, the mapping
    $\zeta\mapsto\Delta(E,\zeta,\varepsilon)$ is
    $\varepsilon$-periodic.
  \end{itemize}
\end{Th}
\noindent Following the now classical definition
(see~\cite{MR1962356,MR2003h:35189,Zw:94,MR2000d:58051}), we set
\begin{Def}
  \label{def:2}
  Fix $\zeta$ real and $E_0$ as in Theorem~\ref{thr:2}.  An energy
  $E\in D(E_0,r_0)$ is a resonance for $H_{\zeta,\varepsilon}$ if it
  is a pole of $E\mapsto R(E,\zeta,\varepsilon)$ i.e. if it is a zero
  of $E\mapsto\Delta(E,\zeta,\varepsilon)$.
\end{Def}
\noindent Our goal is to describe the resonances for
$H_{\zeta,\varepsilon}$ near $E_0$, an energy satisfying the
assumptions of Theorem~\ref{thr:2}. Our main interest is in computing
the imaginary part of the resonances. It is well known that the
existence and the positions of resonances will depend crucially on the
relative position of the spectral window $\mathcal{F}(E):=(E-W)(\R)$
with respect to $\sigma(H_0)$ (see~\cite{MR1962356,MR2003h:35189}).
Therefore, one introduces the set
$\mathcal{W}(E):=(E-W)^{-1}(\sigma(H_{0}))$. It is a subset of the
domain $\mathcal{C}_{C_0}$ where $W$ is analytic.
\subsection{Assumptions on the energy $E_0$}
\label{sec:assumpt-interv-j}
\begin{figure}[htbp]
  \centering \subfigure[]{
    \includegraphics[bbllx=71,bblly=577,bburx=245,bbury=721,width=6cm]{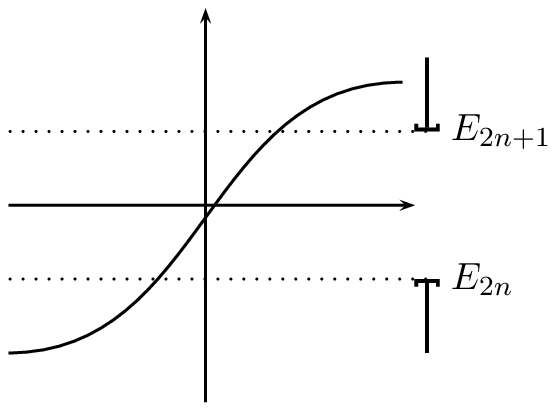}
    \label{fig:5-2}} \hskip1.5cm \subfigure[]{
    \includegraphics[bbllx=71,bblly=577,bburx=245,bbury=721,width=6cm]{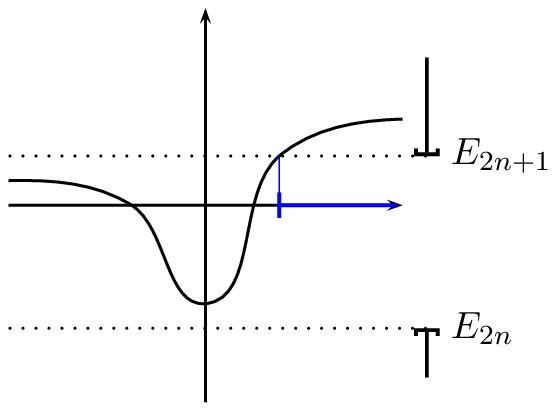}
    \label{fig:6-3}}
  \caption{Examples of $W$ for which assumption (H5) holds.}
  \label{fig:exWH6}
\end{figure}

\begin{figure}[htbp]
  \centering \subfigure[]{
    \includegraphics[bbllx=71,bblly=577,bburx=245,bbury=721,width=6cm]{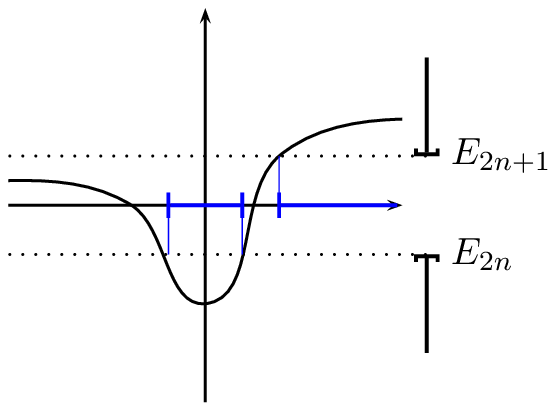}
    \label{fig:6-1}} \hskip1.5cm \subfigure[]{
    \includegraphics[bbllx=71,bblly=577,bburx=245,bbury=721,width=6cm]{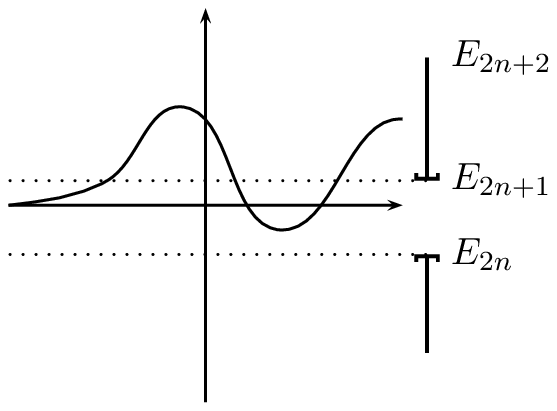}
    \label{fig:6-2}}
  \caption{Examples for which assumption (H6) holds and either $U_-$ or $U_+$ is empty}
  \label{fig:exWH6prim}
\end{figure}

Pick $E_0\in\R$ such that either
$E_0-W_+\in\sigma(H_0)\setminus\partial\sigma(H_0)$, or
$E_0-W_-\in\sigma(H_0)\setminus\partial\sigma(H_0)$, or both hold. A
simple Weyl sequence argument then shows that
$E\in\sigma(H_{\zeta,\varepsilon})$
(for any $\zeta\in\R$ and $\varepsilon>0$).\\
We consider two types of situations and therefore introduce two
different assumptions
\begin{description}
\item [(H5)] the set $\mathcal{W}(E_0)\cap\R$ can be decomposed into
  \begin{equation*}
    \mathcal{W}(E_0)\cap\R=U_{-}(E_0)\cup U_{+}(E_0)
  \end{equation*}
  where $U_{-}(E_0)$ and $U_{+}(E_0)$ are two-by-two disjoint and they
  satisfy:
  \begin{itemize}
  \item
    $U_{-}(E_0)$ and $U_{+}(E_0)$ are either empty or semi-infinite
    intervals, respectively neighborhoods of $-\infty$ and $+\infty$;
  \item the finite edges of $U_{-}(E_0)$ and $U_{+}(E_0)$ (when they
    exist) are not critical points of $W$.
  \end{itemize}
\end{description}
and
\begin{description}
\item [(H6)] the set $\mathcal{W}(E_0)\cap\R$ can be decomposed into
  \begin{equation*}
    \mathcal{W}(E_0)\cap\R=U_{-}(E_0)\cup U(E_0)\cup U_{+}(E_0)
  \end{equation*}
  where $U_{-}(E_0)$, $U(E_0)$ and $U_{+}(E_0)$ are two-by-two
  disjoint and they satisfy:
  \begin{itemize}
  \item $U(E_0)$ is a compact interval not reduced to a single point;
  \item $U_{-}(E_0)$ and $U_{+}(E_0)$ are as in (H5).
  \item the finite edges of $U_{-}(E_0)$, $U_{+}(E_0)$ and $U(E_0)$
    (when they exist) are not critical points of $W$.
  \end{itemize}
\end{description}
%
\begin{figure}[htbp]
  \centering \subfigure[]{
    \includegraphics[bbllx=71,bblly=577,bburx=245,bbury=721,width=6cm]{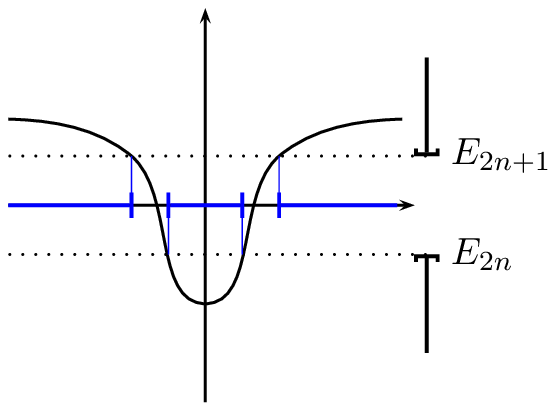}
    \label{fig:5-1}} \hskip1.5cm  \subfigure[]{
    \includegraphics[bbllx=71,bblly=577,bburx=245,bbury=721,width=6cm]{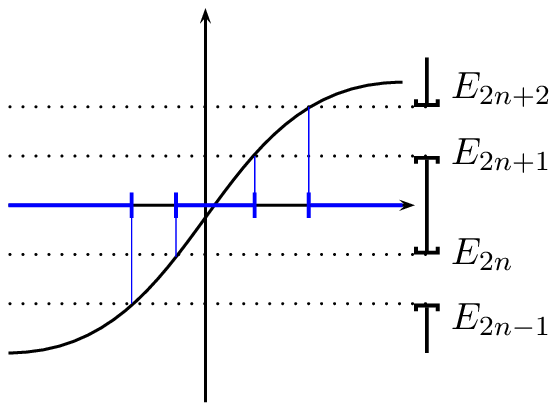}
    \label{fig:5-3}}
  \caption{Examples of $W$ for which assumption (H6) holds and both
    $U_\pm$ are not empty.}
  \label{fig:exWH6sec}
\end{figure}
%
In Fig.~\ref{fig:exWH6},~\ref{fig:exWH6sec} and~\ref{fig:exWH6prim},
we provide some examples of potential profiles; we graphed
$\zeta\mapsto E-W(\zeta)$, and, on the vertical axis represented the
relevant spectral intervals for $H_0$.
%
%
Let us now shortly discuss these assumptions. One easily convinces
oneself that, if (H5) or (H6) holds for some energy $E_0$, it will
hold for all real energies in a neighborhood of $E_0$. Moreover, the
fact that the sets $U_-(E)$ or $U_+(E)$ in the decomposition of
$\mathcal{W}(E)\cap\R$ are empty will not depend on $E$ (in this
neighborhood). Fig.~\ref{fig:exWH6} provides examples where assumption
(H5) holds. Under assumption (H6), Fig.~\ref{fig:exWH6} provides some
examples where both $U_-(E)$ and $U_+(E)$ are non empty, and
Fig.~\ref{fig:exWH6prim}, examples where either $U_-(E)$ or $U_+(E)$
is empty. The two assumptions (H5) and (H6) correspond to the simplest
possible cases: in the general case, i.e. when the sets $E-W(\R)$ and
$\sigma(H_0)$ are in a general position, there can be more than one
compact connected component to the set $\{\zeta\in\R;\ E-W(\zeta)\in
\sigma(H_0)\}$.
%
%
\\ We now state our results on the resonances of $H_{\zeta,\varepsilon}$.
\subsection{Resonance free regions}
\label{sec:reson-free-regi}
We begin our results with the energies a neighborhood of which do not
carry resonances, namely the energies satisfying (H5). Indeed, we
prove
\begin{Th}[\cite{Kl-Ma:05a}]
  \label{thr:3}
  Fix $E_0$ satisfying (H5). There exist $\varepsilon_0>0$,
  $\delta_0>0$, $V_0\subset \C$, a neighborhood of $E_0$ such that,
  for $\varepsilon\in(0,\varepsilon_0)$, for all real $\zeta$, the set
  $V_0$ contains no resonance of $H_{\zeta,\varepsilon}$.
\end{Th}
\noindent This is quite analogous to what is found in the case when
$H_0$ is the free Laplace operator (see
e.g.~\cite{MR871788,Hi-Si:89}).
\subsection{Computing the resonances}
\label{sec:computing-resonances}
To study the real energies close to which one finds resonances, we
introduce a few auxiliary functions necessary to describe the
resonances.
\subsubsection{The complex momentum and its branch points}
\label{sec:complex-momentum-its}
The complex momentum $\zeta\mapsto\kappa(\zeta)$ is defined by
\begin{equation}
  \label{complex-mom}
  \kappa(\zeta)=\kappa(\zeta,E)=k(E-W(\zeta)).
\end{equation}
As $k$, $\kappa$ is analytic and multi-valued. As the branch points of
$k$ are the points $(E_i)_{i\in\N}$, the branch points of $\kappa$
satisfy
\begin{equation}
  \label{BPCM}
  E- W(\zeta)=E_j,\ j\in\N^*.
\end{equation}
As $E$ is real, the set of these points is symmetric with respect to
the real axis, to the imaginary axis and it is $2\pi$-periodic in
$\zeta$. More details are given in section~\ref{sec:kappa}.\\
Fix $E_0$ satisfying assumption (H6). By~\eqref{BPCM}, for $E$
real close to $E_0$, the ends of $U(E)$ and the finite ends of
$U_-(E)$ and $U_+(E)$ (when they are not empty) are branch points of
$\kappa$.  To fix ideas, let us define
\begin{equation*}
  U(E)=[\zeta_0^-(E),\zeta_0^+(E)],\quad
  U_-(E)=(-\infty,\zeta^-(E)]
  \quad\text{and}\quad U_+(E)=[\zeta^+(E),+\infty)
\end{equation*}
When $U_+(E)$ (resp. $U_-(E)$) is empty, we set $\zeta^+(E)=+\infty$
(resp. $\zeta^-(E)=-\infty$).\\
One shows that there exists a determination of the complex momentum,
say $\kappa_0$, and a real neighborhood of $E_0$, say $V_0^\R$, such
that, for $E\in V_0^\R$, $\kappa_0(\zeta)\in[0,\pi]$ for $\zeta\in U(E)$.
One defines
\begin{equation}
  \label{eq:5}
  \Phi_0(E)=\int_{\zeta_0^-(E)}^{\zeta_0^+(E)}\kappa_0(\zeta,E)d\zeta
  \quad\text{and}\quad\delta\kappa=\frac1\pi[\kappa(\zeta_0^+(E),E)
  -\kappa_0(\zeta_0^-(E),E)].
\end{equation}
One shows
\begin{Le}[\cite{Kl-Ma:05a}]
  \label{le:1}
  The constant $\delta\kappa$ belongs to $\{-1,1,0\}$
  and is independent of $E$ in $V_0^\R$. The function $\Phi_0:\
  V_0^\R\to\R^+$ can be extended to a function analytic in a complex
  neighborhood of $E_0$.  Moreover, there exists $C>0$ such that, in
  this neighborhood, one has
  \begin{equation}
    \label{eq:10}
    |\Phi'_0(E)|\geq\frac1C \quad\text{and}\quad\Phi'_0(E)\cdot
    \delta\kappa\geq0.
  \end{equation}
\end{Le}
\noindent Assume $U_-(E)$ or $U_+(E)$ or both are not empty. One then
proves that the imaginary part of the determination $\kappa_0$ keeps a
fixed sign in the interval between $U_-(E)$ and $U(E)$, and between
$U(E)$ and $U_+(E)$, i.e. on the intervals $(\zeta_-(E),\zeta_
0^-(E))$ and $(\zeta_0^+(E),\zeta_+(E))$. We define
\begin{equation}
  \label{eq:6}
  S_-(E)=\pm\int_{\zeta_-(E)}^{\zeta_ 0^-(E)}\text{Im}\,\kappa(\zeta)d\zeta
  \quad\left(\text{resp.}\quad
    S_+(E)=\pm\int_{\zeta_0^+(E)}^{\zeta_+(E)}\text{Im}\,\kappa(\zeta)d\zeta
  \right)
\end{equation}
where $\pm$ is chosen so that $S_+(E)$ and $S_-(E)$ are positive. One
proves
\begin{Le}[\cite{Kl-Ma:05a}]
  \label{le:2}
  Pick $\nu\in\{+,-\}$. The function $S_\nu:\ V_0^\R\to\R^+$ can be
  extended to a function analytic in a complex neighborhood of $E_0$.
\end{Le}
\noindent When $U_+(E_0)$ (resp. $U_-(E_0)$) is empty, it stays
empty for $E$ close to $E_0$, and one sets $S_+(E)=+\infty$ (resp.
$S_-(E)=+\infty$).  Define the tunneling coefficients
\begin{equation}
  \label{eq:7}
  t_{\pm}(E)=e^{-S_\pm(E)/\varepsilon}\quad\text{and}\quad t(E)=t_+(E)+t_-(E).
\end{equation}
\subsubsection{Resonances}
\label{sec:resonances}
We prove
\begin{Th}[\cite{Kl-Ma:05a}]
  \label{thr:4}
  Fix $E_0$ satisfying (H6). There exist $\varepsilon_0>0$,
  $\delta_0>0$, $V_0\subset \C$, a neighborhood of $E_0$, and a real
  analytic function $E\mapsto\check \Phi(E,\varepsilon)$ defined on
  $V_0\times[0,\varepsilon_0]$ satisfying the uniform asymptotics
  \begin{equation}
    \label{eq:17}
    \check\Phi(E,\varepsilon)=\Phi(E)+O(\varepsilon)\quad\text{when
    }\varepsilon\to0,
  \end{equation}
  such that, if one defines the finite sequence of points in
  $V_0\cap\R$, say $(E^l)_l:=(E^l(\zeta,\varepsilon))_l$, by
  \begin{equation}
    \label{eq:11}
    \frac1{\varepsilon}\check\Phi(E^l,\varepsilon)=
    \frac1{\varepsilon}\pi\,\delta\kappa\,\zeta
    +\frac\pi2+\pi l,\quad l\in\Z,
  \end{equation}
  then, for $\varepsilon\in(0,\varepsilon_0)$, for all real $\zeta$,
  the resonances of $H_{\zeta,\varepsilon}$ in $V_0$ are contained in
  the union of the disks
  \begin{equation}
    \label{eq:8}
    D^l:=\left\{E\in V_0;\ |E-E^l|\leq e^{-\delta_0/\varepsilon}\right\}.
  \end{equation}
  More precisely, each of these disks contains exactly one simple
  resonance, say $\tilde E^l(\zeta,E)$, that satisfies
  \begin{equation}
    \label{eq:9}
    \text{Im}\,(\tilde E^l(\zeta,E))=\varepsilon\cdot c_0\cdot t(E^l)(1+o(1))
  \end{equation}
  where $o(1)\to0$ when $\varepsilon\to0$ and $c_0$ is a constant.
\end{Th}
\noindent Let us discuss the resonances obtained in
Theorem~\ref{thr:4}, in particular, their behavior as functions of
$\zeta$. By Theorem~\ref{thr:2}, the function
$\zeta\mapsto\Delta(E,\zeta,\varepsilon)$ is
$\varepsilon$-periodic and by Theorem~\ref{thr:4}, the resonances
are simple, the resonances being the zeroes of
$E\mapsto\Delta(E,\zeta,\varepsilon)$, they are
$\varepsilon$-periodic functions of $\zeta$.
\par Theorem~\ref{thr:4} says that their imaginary part does not
oscillate to first order. It also gives some information about the
oscillations of the real part. Therefore one has to distinguish
the two cases  $\delta\kappa\not=0$ and  $\delta\kappa=0$.
\par When $\delta\kappa\not=0$, the quantization
condition~\eqref{eq:11} and equations~\eqref{eq:10}
and~\eqref{eq:11} show that the real part of the resonances is
monotonous in $\zeta$. But, as they come in a sequence,
renumbering the resonances, they can also be seen as oscillating
with amplitude roughly $C\varepsilon$ (for $C>0$). Of course, to
see this oscillation phenomenon, one needs to renumber the
resonances. The situation is similar to that encountered when
studying Stark-Wannier resonances (see~\cite{MR99d:34164}); in
this case also, the resonances still exhibit oscillations of
frequency $2\pi/\varepsilon$ but the amplitude is $\varepsilon$
(the Stark-Wannier ladders).
\par When $\delta\kappa\not=0$, Theorem~\ref{thr:4} shows that the
oscillations of the real part are at most exponentially small. Under
more precise assumption than those made in the present talk, one can
compute the amplitude of the oscillations of the resonances
(\cite{Kl-Ma:05b}). An analogous behavior was found for the
eigenvalues of $H_{\zeta,\varepsilon}$ outside the essential spectrum
in~\cite{Mar:04}; they oscillate with a frequency $2\pi/\varepsilon$
and an amplitude that is exponentially small in $\varepsilon$; the
amplitude was given by a complex tunneling coefficient.
\par Moving the energy $E$ (for a fixed $W$), one sees that, in some
cases, one can pass continuously from assumption (H5) to (H6) and
vice-versa.  It would be interesting to study what happens to the
resonances uncovered in Theorem~\ref{thr:4} when one crosses this
transition. Such a study has been done in~\cite{MR1635811,MR1972758}
in the case when $V\equiv0$.
\subsection{The heuristics}
\label{sec:heuristics}
We now discuss the heuristics explaining these results. In the figures
below, in the part indexed a), we represented the local configurations
of the potential to show the assumption made on the energy $E$ in the
different cases. In the part indexed b), we represented the phase
space picture of the iso-energy curve i.e. the sets
$\{(\kappa,\zeta)\in\R^2;\ \kappa=\kappa(E-W(\zeta))\}$. It is
$2\pi$-periodic in the $\kappa$-direction (i.e. the vertical
direction) so we depicted only two periods.
%
\begin{figure}[htbp]
  \centering
  \subfigure[]{
    \includegraphics[bbllx=71,bblly=577,bburx=245,bbury=721,width=6cm]{DESSINS/exple5-1.eps}
    \label{fig:pot1-1}}
  \hskip1.5cm
  \subfigure[]{
    \includegraphics[bbllx=0,bblly=0,bburx=544,bbury=442,width=6cm]{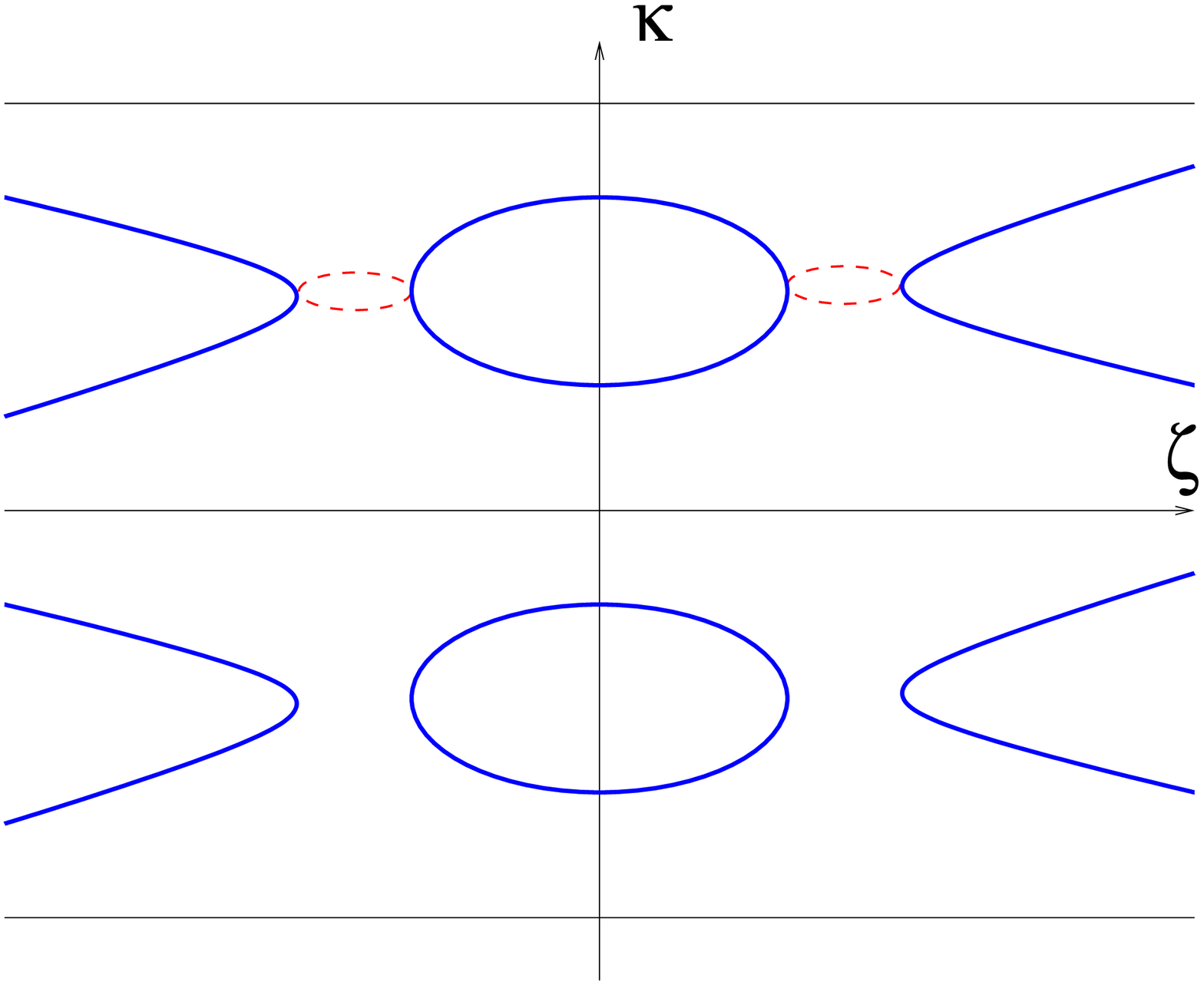}
    \label{fig:esp1-1}}
  \caption{The phase space picture}
  \label{fig:esp1}
\end{figure}
%
On the same picture, we also show some loops that are dashed. These
are special loops in $\{(\kappa,\zeta)\in\C^2;\
\kappa=\kappa(E-W(\zeta))\}$ that join the connected components of
$\{(\kappa,\zeta)\in\R^2;\ \kappa=\kappa(E-W(\zeta))\}$.\\
Assumption (H5) just means that the real iso-energy curve is empty, in
which case there are no resonances as we saw. We now assume (H6)
holds.
\par In Fig.~\ref{fig:esp1}, the compact connected components of the
iso-energy curve consist of a single torus per periodicity cell.
The torus gives rise to the quantization condition~\eqref{eq:17}:
the corresponding phase is obtained (to first order) by
integrating the canonical one-form $\kappa d\zeta$ along this
torus. The fact that this corresponds to a resonance rather than
to a eigenvalue is due to the fact that the iso-energy curve has
components leading to infinity. The dashed curve represents the
instanton linking the two. To compute the action determining the
lifetime of the resonance, one integrates the canonical one-form
$\kappa d\zeta$ along this curve.
%
\begin{figure}[htbp]
  \centering
  \subfigure[]{
    \includegraphics[bbllx=71,bblly=577,bburx=245,bbury=721,width=6cm]{DESSINS/exple5-3.eps}
    \label{fig:pot1-2}}
  \hskip1.5cm
  \subfigure[]{
    \includegraphics[bbllx=0,bblly=0,bburx=544,bbury=442,width=6cm]{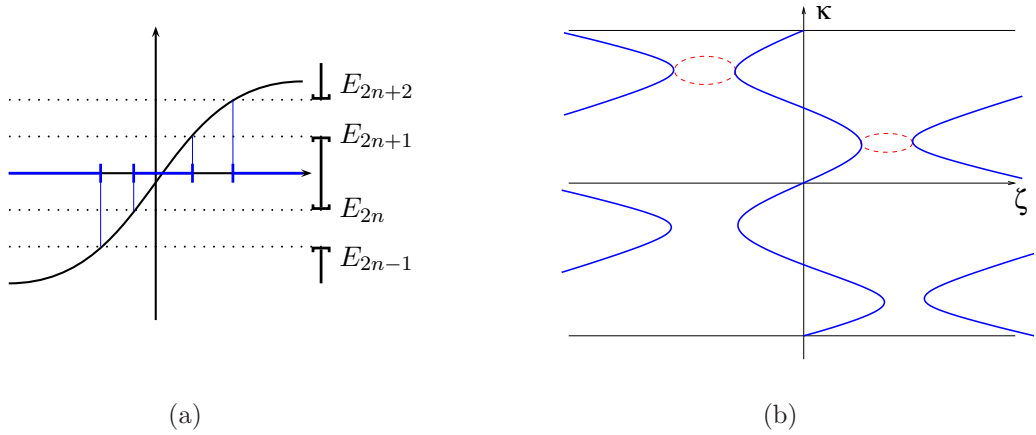}
    \label{fig:esp1-2}}
  \caption{The phase space picture}
  \label{fig:esp2}
\end{figure}
%
\par In Fig.~\ref{fig:esp2}, the compact connected components of the
iso-energy curve are absent. Nevertheless, one can compute a phase (to
first order) by integrating the canonical one-form $\kappa d\zeta$
along a period of the central curve i.e. the one that is extended in
the $\kappa$-direction. When the connected component extended in the
$\zeta$-directions are absent, one can show that such a phase does not
give rise to eigenvalues. The presence of these components leading to
infinity in the $\zeta$-direction gives rise to resonances.  Again, to
compute the action determining the lifetime of the resonance, one
integrates the canonical one-form $\kappa d\zeta$ along the dashed
curve.
\section{The adiabatic complex WKB method and the proof of
Theorem~\ref{thr:2}}
\label{sec:complex-wkb-method}
In~\cite{MR2002h:81069} and~\cite{MR2097997}, we have developed a new
asymptotic method to study solutions to an adiabatically perturbed
periodic Schr{\"o}dinger equation i.e., to study solutions of the equation
\begin{equation}
  \label{G.2a}
 -\frac{d^2}{dx^2}\psi(x,\zeta)+(V(x)+W(\varepsilon
  x+\zeta))\psi(x,\zeta)= E\psi(x,\zeta)
\end{equation}
in the limit $\varepsilon\to 0$. The function $\zeta\mapsto W(\zeta)$
is an analytic function in a neighborhood of the real axis. The main
idea of the method is to get the information on the behavior of the
solutions in $x$ from the study of their behavior on the complex plane
of $\zeta$. The natural condition allowing to relate the behavior in
$\zeta$ to the behavior in $x$ is the consistency condition
\begin{equation}
  \label{consistency:1}
  \psi(x+1,\,\zeta,E)=\psi(x,\,\zeta+\varepsilon,E).
\end{equation}
One can construct solutions to both~\eqref{G.2a}
and~\eqref{consistency:1} that have a simple asymptotic behavior on
certain domains of the complex plane of $\zeta$.\\
The ideas underlying this result are simple. For solutions of
\eqref{G.2a} satisfying~\eqref{consistency:1}, it is equivalent to
know their behavior for large $x$ or for large $\zeta$. Hence, the
idea is to study their behavior for large $\zeta$. Keeping $x$ fixed
to some compact interval, say $[-1,1]$, to first order in
$\varepsilon$, a solution to~\eqref{G.2a} is a solution to
\begin{equation}\label{PSE}
  -\frac{d^2}{dx^2}\psi\,(x)+ V\,(x)\psi\,(x)=\mathcal{E}\psi\,(x),
   \quad x\in\R,
\end{equation}
for $\mathcal{E}=E-W(\zeta)$. Hence, the Bloch solutions are special
solutions to this equation. From these solution, multiplying them by a
suitably chosen function of $\zeta$ (i.e. independent of $x$), one can
construct solution to both~\eqref{G.2a} and~\eqref{consistency:1} at
the same time. The standard form of the solutions is given below
in~\eqref{eq:2}.
\par We first describe the standard asymptotic behavior of consistent
solutions; in order to do this, we recall some fact on periodic
Schr{\"o}dinger operators on $\R$. Then, we review the theory developed
in~\cite{MR2002h:81069,MR2097997} on the existence of consistent
solutions with these asymptotics. We next describe some new results
used to control the behavior of the solutions at infinity. We conclude
this section with the proof of Theorem~\ref{thr:2}.
\subsection{Periodic Schr{\"o}dinger operators}
\label{S3}
\noindent In this section, we discuss the periodic Schr{\"o}din\-ger
operator~\eqref{Ho} where $V$ is a $1$-periodic, real valued,
$L^2_{loc}$-function. First, we collect well known results needed in
the present paper
(see~\cite{MR2002f:81151,Eas:73,Ma-Os:75,MR80b:30039,Ti:58}). In the
second part of the section, we introduce a meromorphic differential
defined on the Riemann surface associated to the periodic operator.
This object plays an important role for the adiabatic constructions
(see~\cite{MR2097997}).
\subsubsection{Bloch solutions}
\label{sec:bloch-solutions}
Let $\psi$ be a nontrivial solution of the equation~\eqref{PSE}
satisfying the relation $\psi\,(x+1)=\lambda\,\psi\,(x)$ for all
$x\in\R$ with $\lambda\in\C$ independent of $x$. Such a solution is
called a {\it Bloch solution}, and the number $\lambda$ is called the
{\it Floquet multiplier}. We now discuss properties of Bloch solutions
(see~\cite{MR2002f:81151}).
\smallpagebreak As in section~\ref{sec:periodic-operator}, we denote
the spectral bands of the periodic Schr{\"o}dinger equation by
$[E_1,\,E_2]$, $[E_3,\,E_4]$, $\dots$, $[E_{2n+1},\,E_{2n+2}]$,
$\dots$. Consider $\mathcal{S}_\pm $, two copies of the complex plane
$\mathcal{E}\in\C$ cut along the spectral bands. Paste them together
to get a Riemann surface with square root branch points.  We denote
this Riemann surface by $\mathcal{S}$. In the sequel, $\pi_c:\
{\mathcal S}\mapsto\C$ is the canonical projection.
\smallpagebreak One can construct a Bloch solution
$\psi(x,\mathcal{E})$ of equation~\eqref{PSE} meromorphic on $\mathcal
S$.  For any $\mathcal{E}$, we normalize it by the condition
$\psi(1,\mathcal{E})= 1$. Then, the poles of
$\mathcal{E}\mapsto\psi(x,\mathcal{E})$ are projected by $\pi_c$
either in the open spectral gaps or at their ends. More precisely,
there is exactly one simple pole per open gap.  The position of the
pole is independent of $x$ (see~\cite{MR2002f:81151}).
\smallpagebreak Let $\hat{\cdot}:\ {\mathcal S}\mapsto {\mathcal S}$
be the canonical transposition mapping: for any point
$\mathcal{E}\in\mathcal{S}$, the point $\hat{\mathcal{E}}$ is the
unique solution to the equation $\pi_c(\mathcal{E})=E$ different
from $\mathcal E$ outside the branch points.\\
The function $x\mapsto \psi(x,{\hat{\mathcal E}})$ is one more Bloch
solution of~\eqref{PSE}. Except at the edges of the spectrum (i.e. the
branch points of $\mathcal{S}$), the functions $\psi(\cdot,{\mathcal
  E})$ and $\psi(\cdot,\hat {\mathcal E})$ are linearly independent
solutions of~\eqref{PSE}. In the spectral gaps, they are real valued
functions of $x$, and, on the spectral bands, they differ only by the
complex conjugation (see~\cite{MR2002f:81151}).
\subsubsection{The Bloch quasi-momentum}
\label{SS3.2}
Consider the Bloch solution $\psi(x,\mathcal{E})$. The corresponding
Floquet multiplier $\lambda\,(\mathcal{E})$ is analytic on
$\mathcal{S}$. Represent it in the form
$\lambda(\mathcal{E})=\exp(ik(\mathcal{E}))$. The function
$k(\mathcal{E})$ is the {\it Bloch quasi-momentum}.
\\ The Bloch quasi-momentum is an analytic multi-valued function of
$\mathcal{E}$. It has the same branch points as $\psi(x,\mathcal{E})$
(see~\cite{MR2002f:81151}).
\\ Let $D\in \C$ be a simply connected domain containing no branch
point of the Bloch quasi-momentum $k$. On $D$, fix $k_0$, a continuous
(hence, analytic) branch of $k$. All other branches of $k$ that are
continuous on $D$ are then given by the formula
\begin{equation*}
  \label{eq:55}
   k_{\pm ,l}({\mathcal E})=\pm k_0({\mathcal E})+2\pi l,\quad l\in\Z.
\end{equation*}
All the branch points of the Bloch quasi-momentum are of square root
type: let $E_l$ be a branch point, then, in a sufficiently small
neighborhood of $E_l$, the quasi-momentum is analytic as a function of
the variable $\sqrt{\mathcal{E}-E_l}$; for any analytic branch of $k$,
one has
\begin{equation*}
  \label{sqrt}
  k(\mathcal{E})=k_l+c_l\sqrt{\mathcal{E}-E_l}+O(\mathcal{E}-E_l),\quad c_l\not=0,
\end{equation*}
with constants $k_l$ and $c_l$ depending on the branch.\\
Let $\C_+$ be the upper complex half-plane. There exists $k_p$, an
analytic branch of $k$ that conformally maps $\C_+$ onto the quadrant
$\{k\in\C;\ \im k> 0,\,\,\re k> 0\}$ cut along compact vertical
intervals, say $\pi l+i I_l$ where $l\in\N^*$ and $I_l\subset\R$,
(see~\cite{MR2002f:81151}). The branch $k_p$ is continuous up to the
real line.  It is real and increasing along the spectrum of $H_0$; it
maps the spectral band $[E_{2n-1}, E_{2n}]$ on the interval
$[\pi(n-1),\pi n]$. On the open gaps, $\re k_p$ is constant, and $\im
k_p$ is positive and has exactly one maximum; this maximum is non
degenerate.\\
We call $k_p$ the {\it main} branch of the Bloch
quasi-momentum.\\
Finally, we note that the main branch can be analytically continued on
the complex plane cut only along the spectral gaps of the periodic
operator.
\subsubsection{Meromorphic differential $\Omega$}
\label{sec:Omega}
On the Riemann surface $\mathcal S$,
consider the function
\begin{equation}\label{omega}
\omega({\mathcal E})=
-\frac{\int_0^1 \psi(x,\hat {\mathcal E})\,\left(\dot\psi(x,{\mathcal
      E})-i\dot k({\mathcal E}) x\,\psi(x,{\mathcal E})\right)\,dx}
{\int_0^1\psi(x,{\mathcal E})\psi(x,\hat {\mathcal E}) dx}.
\end{equation}
where $k$ is the Bloch quasi-momentum of $\psi$,
and the dot denotes the partial derivative with respect to ${\mathcal E}$. This
function was introduced in~\cite{MR2156718} (the definition given in
that paper is equivalent to~\eqref{omega}). In~\cite{MR2156718}, we
have proved that $\omega$ has the following properties:
\begin{enumerate}
\item the differential $\Omega=\omega\,d{\mathcal E}$ is meromorphic
  on $\mathcal S$; its poles are the points of $P\cup Q$, where $P$ is
  the set of poles of $\psi(x,{\mathcal E})$, and $Q$ is the set of
  points where $k'({\mathcal E})=0$;
\item all the poles of $\Omega$ are simple;
\item $\forall p\in P\setminus Q$, $\res_p \Omega=1$; $\forall q\in
  Q\setminus P$, $\res_q\Omega=-1/2$; $\forall r\in P\cap Q$,
  $\res_r\Omega=1/2$.
\item if $\pi_c({\mathcal E})$ belongs to a gap, then
  $\omega({\mathcal E})\in \R$;
\item if $\pi_c({\mathcal E})$ belongs to a band, then
  $\overline{\omega({\mathcal E})}=\omega(\hat {\mathcal E})$.
\end{enumerate}
\subsection{Standard behavior of consistent solutions}
\label{sec:stand-behav-solut}
We now discuss in more detail two analytic objects central to the
complex WKB method, the complex momentum defined
in~\eqref{complex-mom} and the canonical Bloch solutions defined
below. For $\zeta\in\mathcal{D}(W)$, the domain of analyticity of the
function $W$, we define
\begin{equation}
  \label{eq:29}
  {\mathcal E}(\zeta)=E-W(\zeta)
\end{equation}
The complex momentum and the canonical Bloch solutions are the Bloch
quasi-momentum and particular Bloch solutions of the equation
\begin{equation}
  \label{eq:52}
  -\frac{d^2}{dx^2}\psi+V\psi={\mathcal E}(\zeta)\psi.
\end{equation}
considered as functions of $\zeta$.
\subsubsection{The complex momentum}
\label{sec:kappa}
For $\zeta\in\mathcal{D}(W)$, the domain of analyticity of the
function $W$, the complex momentum is given by the formula
$\kappa(\zeta)=k(\mathcal{E}(\zeta))$ where $k$ is the Bloch
quasi-momentum of~\eqref{Ho}. Clearly, $\kappa$ is a multi-valued
analytic function; a point $\zeta$ such that $W'(\zeta)\ne0$ is a
branch point of $\kappa$ if and only if it satisfies~\eqref{BPCM}.
All the branch points of the complex momentum are of square root type.
\smallpagebreak A simply connected set $D\subset\mathcal{D}(W)$
containing no branch points of $\kappa$ is called {\it regular}. Let
$\kappa_p$ be a branch of the complex momentum analytic in a regular
domain $D$. All the other branches analytic in $D$ are described by
\begin{equation}
  \label{allbr}
  \kappa_m^\pm  =\pm  \kappa_p+2\pi m\quad\text{where}\quad m\in\Z.
\end{equation}
\subsubsection{Canonical Bloch solutions}
\label{sec:canon-bloch-solut}
To describe the asymptotic formulae of the complex WKB method, one
needs Bloch solutions of equation~\eqref{eq:52} analytic in $\zeta$ on
a given regular domain. We build them using the 1-form
$\Omega=\omega\,d\mathcal{E}$ introduced in section~\ref{sec:Omega}.
\smallpagebreak Pick $\zeta_0$, a regular point i.e. a point that is
not a branch point of $\kappa$. Let
$\mathcal{E}_0=\mathcal{E}(\zeta_0)$. Assume that
$\mathcal{E}_0\not\in P\cup Q$ (the sets $P$ and $Q$ are defined in
section~\ref{sec:Omega}). In $U_0$, a sufficiently small neighborhood
of $\mathcal{E}_0$, we fix $k$, a branch of the Bloch quasi-momentum,
and $\psi_\pm(x,\mathcal{E})$, two branches of the Bloch solution
$\psi(x,\mathcal{E})$ such that $k$ is the Bloch quasi-momentum of
$\psi_+$.  Also, in $U_0$, consider $\Omega_\pm$, the two
corresponding branches of $\Omega$, and fix a branch of the function
$\mathcal{E}\mapsto q(\mathcal{E})=\sqrt{k'(\mathcal{E})}$.  Assume
that $V_0$ is a neighborhood of $\zeta_0$ such that ${\mathcal
  E}(V_0)\subset U_0$.  For $\zeta\in V_0$, we let
\begin{equation}
  \label{canonicalBS}
  \Psi_\pm (x,\zeta)=
  q(\mathcal{E})\,e^{\int_{\mathcal{E}_0}^\mathcal{E} \Omega_\pm}
  \psi_\pm (x,\mathcal{E}),\quad\text{where}\quad
  \mathcal{E}=\mathcal{E}(\zeta).
\end{equation}
The functions $\Psi_\pm$ are the {\it canonical Bloch solutions
  normalized at the point $\zeta_0$}. The quasi-momentum associated to
these solutions is $\kappa(\zeta)=k(E-W(\zeta))$.
\smallpagebreak The properties of the differential $\Omega$ imply that
the solutions $\Psi_\pm$ can be analytically continued from
$V_0$ to any regular domain $D$ containing $V_0$. \\
One has (see~\cite{MR2002h:81069})
\begin{equation}
  \label{Wcanonical}
  w(\Psi_+(\cdot ,\zeta),\Psi_-(\cdot ,\zeta))=w(\Psi_+(\cdot ,\zeta_0),\Psi_-(\cdot ,\zeta_0))=
  k'(\mathcal{E}_0)w(\psi_+(\cdot,\mathcal{E}_0),\psi_-(\cdot,\mathcal{E}_0))
\end{equation}
As $\mathcal{E}_0\not\in Q\cup\{E_l,\ l\geq1\}$, the Wronskian
$w(\Psi_+(\cdot ,\zeta),\Psi_-(\cdot ,\zeta))$ does not vanish.
\subsubsection{Solutions having standard asymptotic behavior}
\label{sec:standard-asymptotics}
Fix $E=E_0$. Let $D$ be a regular domain. Fix $\zeta_0\in D$ so that
$\mathcal{E}(\zeta_0)\not\in P\cup Q$. Let $\kappa$ be a branch of the
complex momentum continuous in $D$, and let $\Psi_\pm$ be the
canonical Bloch solutions defined on $D$, normalized at $\zeta_0$ and
indexed so that $\kappa$ be the quasi-momentum for $\Psi_+$.\\
We recall the following basic definition from~\cite{MR2097997}
\begin{Def}
  \label{def:1}
  Fix $\eta\in\{+,-\}$, $\zeta_0\in D$, $X>1$, $V_0$, a complex
  neighborhood of some energy $E_0$ and $D$ a domain in the
  $\zeta$-plane. We say that $f$, a solution of~\eqref{G.2a}, has
  standard asymptotics (or standard behavior) $f\sim\exp(\eta\frac{i}
  {\varepsilon} \int_{\zeta_0}^{\zeta}\kappa\,du)\cdot\Psi_\eta$ in
  $(-X,X)\times D\times V_0$ if
  \begin{itemize}
  \item $f$ is defined and satisfies~\eqref{G.2a}
    and~\eqref{consistency:1} for any $(x,\zeta,E)\in (-X,X)\times
    D\times V_0$;
  \item $f$ is analytic in $\zeta\in D$ and in $E\in V_0$;
  \item for any $K$, compact subset of $D$, there exists $V\subset
    V_0$, a neighborhood of $E_0$, such that, for $(x,\zeta,E)\in
    (-X,X)\times K\times V$, $f$ has the uniform asymptotics
    \begin{gather*}
      f(x,\zeta,E,\varepsilon)=e^{\dsize \eta\,\frac{i}{\varepsilon}
        \int_{\zeta_0}^{\zeta}
        \kappa(u,E)\,du}\, (\Psi_\eta(x,\zeta,E)
      +g_\eta(x,\zeta,E,\varepsilon))\\
      \intertext{where}\lim_{\varepsilon\to0}\
      \sup_{\substack{x\in(-X,X)\\\zeta\in K\\ E\in V}}
      |g_\eta(x,\zeta,E,\varepsilon)|=0.
    \end{gather*}
  \item this asymptotic can be differentiated once in $x$.
\end{itemize}
\end{Def}
\noindent Let $(f_+,f_-)$ be two solutions of~\eqref{G.2a} having
standard behavior $f_\pm\sim e^{\dsize\pm\frac{i}{\varepsilon}
  \int^{\zeta} \kappa\,d\zeta}\,\Psi_\pm$ in $(-X,X)\times D\times V_0$. One
computes
\begin{equation*}
  w(f_+,f_-)=w(\Psi_+,\Psi_-)+o(1).
\end{equation*}
By~\eqref{Wcanonical}, for $\zeta$ in any fixed compact subset of $D$
and $\varepsilon$ sufficiently small, the solutions $(f_+,f_-)$ are
linearly independent.
\subsection{The main theorem of the adiabatic complex WKB method }
\label{sec:main-theor-adiabt}
A basic and important example of a domain where one can construct a
solution with standard asymptotic behavior is a canonical domain. Let
us define canonical domains and formulate the basic result of the
adiabatic complex WKB method.
\subsubsection{Canonical lines}
\label{sec:canonical-lines}
We recall that a curve $\gamma$ is vertical if it intersects the lines
$\{\im \zeta=\Const\}$ at non-zero angles $\theta$, \ $0<\theta<\pi$.
Vertical lines are naturally parametrized by $\im\zeta$.\\
A set $D$ in the $\zeta$-plane is said to be bounded regular at energy
$E_0$ if $\overline{D}$, the closure of $D$, is compact and does not
contain any branch point of $\zeta\mapsto\kappa(\zeta,E_0)$.
\smallpagebreak Let $\gamma$ be a $C^1$ bounded regular vertical curve
at energy $E_0$. On $\gamma$, fix a point $\zeta_0$ and $\kappa$, a
continuous branch of the complex momentum.
\begin{Def}
  \label{def:3}
  The curve $\gamma$ is {\it canonical} with respect to $\kappa$ at
  energy $E_0$ if, along $\gamma$, for $y=\im\zeta$, one has
  \begin{equation*}
    \frac d{dy}\,\im\left(\int_{\zeta_0}^\zeta\kappa(u,E_0)du\right)>0
    \quad\text{and} \quad \frac d{dy}\,\left(
    \im\int_{\zeta_0}^\zeta(\kappa(u,E_0)-\pi)du\right)<0
  \end{equation*}
\end{Def}
\noindent One easily checks that,
\begin{itemize}
\item if a vertical curve is bounded regular at energy $E_0$, it is bounded regular at any energy $E$ in a neighborhood of $E_0$;
\item if a vertical curve is bounded regular and canonical with
  respect to $\kappa$ at energy $E_0$, it is also canonical with
  respect to $\kappa$ at any energy $E$ in a neighborhood of $E_0$.
\end{itemize}
\subsubsection{Canonical domains}
\label{sec:canonical-domains}
Let $K$ be a bounded regular domain at energy $E_0$ i.e.
$\overline{K}$ is compact and does not contain any branch point of
$\zeta\mapsto\kappa(\zeta,E_0)$. On $K$, fix a continuous branch of
the complex momentum, say $\kappa$.
\begin{Def}
  \label{def:4}
  The domain $K$ is called {\it canonical} (with respect to $\kappa$
  at energy $E_0$) if there exists two points $\zeta_1$ and $\zeta_2$
  located on $\partial K$, the boundary of $K$, such that $K$ is the
  union of curves connecting $\zeta_1$ and $\zeta_2$ that are
  canonical (with respect to $\kappa$ at energy $E_0$).
\end{Def}
\subsubsection{The main theorem of the adiabatic complex WKB theory}
\label{sec:main-theorem}
One has
\begin{Th}[\cite{MR2002h:81069,MR2097997}]
  \label{T5.1}
  Let $K$ be a bounded domain canonical with respect to $\kappa$ at
  energy $E_0$. Fix $X>1$ and $\zeta_0\in K$. Then, there exists
  $V_0$, a neighborhood of $E_0$ such that, for sufficiently small
  positive $\varepsilon$, there exist $(f_\pm)$, two solutions of
  \eqref{family}, having the standard behavior in $(-X,X)\times
  K\times V_0$ that is
  \begin{equation*}
    f_\pm \sim \exp\left(\pm \frac{i}{\varepsilon}
      \int_{\zeta_{0}}^{\zeta} \kappa d\zeta\right)\Psi_\pm.
  \end{equation*}
  For any fixed $x\in\R$, the functions $(\zeta,E)\mapsto f_\pm
  (x,\zeta,E)$ are analytic in $S(K)\times V_0$ where $S(K)$ is the
  smallest horizontal strip containing $K$.
\end{Th}
\noindent In addition to this result, we need analogues for
domains that are neighborhoods of infinity. Such results were
developed in~\cite{Mar:02,Mar:04}; they are not sufficient for the
purpose of the present paper. Hence, we present two new results in the
next section.
\subsection{Consistent solutions at infinity}
\label{sec:consistent-basis-at}
The standard behavior at infinity is given by
\begin{Th}[\cite{Kl-Ma:05a}]
  \label{thr:1}
  Assume (H1)-(H4) are satisfied. Fix $\eta\in\{+,-\}$ and
  $\alpha\in(0,1)$. Pick $E_0\in\R$ such that either
  $E_0-W_\eta\in\sigma(H_0)\setminus\partial\sigma(H_0)$. Fix $X>1$
  and $C_1>C_0$ (where $C_0$ is defined in assumption (H4),
  section~\ref{sec:assumptions}).  Then, there exist $C>0$ and
  $\delta>0$ such that, for any $E\in D(E_0, \delta)$, the cone
  $\mathcal{C}_\eta=\{\zeta\in\C;\ \eta\text{Re}\,\zeta>C,\ C_1
  |\text{Im}\,\zeta|<\eta\text{Re}\,\zeta\}$ is a canonical domain.
  More precisely, fix $\zeta_0\in\mathcal{C}_\eta\cap\R$. Then, there
  exist a branch of the quasi-momentum
  $\kappa(\zeta)=\kappa(\zeta,E)$ that satisfies
  \begin{equation}
    \label{eq:1}
    \forall(\zeta,E)\in\mathcal{C}_\eta\times D(E_0,\delta),\quad
    0<\text{Re}\,\kappa(\zeta,E)<\pi
  \end{equation}
  and a function $(x,\zeta,E)\mapsto f(x,\zeta,E)$ defined on
  $(-X,X)\times\mathcal{C}_\eta\times D(E_0,\delta)$ that satisfies:
  \begin{itemize}
  \item it is a consistent solution to equation~\eqref{G.2a}
  \item for $x\in(-X,X)$, $(\zeta,E)\mapsto f(x,\zeta,E)$ is
    analytic on $\mathcal{C}_\eta\times D(E_0,\delta)$;
  \item the function $f$ admits the asymptotics
    \begin{equation}
      \label{eq:2}
       f(x,\zeta,E)=e^{\frac{i}{\varepsilon}
        \int_{\zeta_0}^{\zeta}\kappa(u,E)du}
      \left(\Psi(x,\zeta,E)+g(x,\zeta,E,\varepsilon)\right)
    \end{equation}
    where $\Psi$ is the canonical Bloch solutions $\Psi_+$ associated
    to $\kappa$ (see~\eqref{canonicalBS}) and
    \begin{equation}
      \label{eq:47}
      \sup_{\varepsilon\in(0,\varepsilon_0)}
      \sup_{\substack{\zeta\in\mathcal{C}_+\\E\in
          D(E_0,\delta)}}|\varepsilon^{\alpha-1}
      \zeta^sg(x,\zeta,E,\varepsilon)|<+\infty.
    \end{equation}
  \item the asymptotics can be differentiated once in $x$;
  \item define the function
    $(x,\zeta,E)\mapsto f^*(x,\zeta,E)=\overline{ f(\overline{x},
      \overline{\zeta},\overline{E})}$ ; then, $( f, f^*)$ form a
    basis of consistent solutions and satisfy
    \begin{equation}
      \label{eq:3}
      w( f(\cdot ,\zeta,E), f^*(\cdot ,\zeta,E))=
      w(\Psi_+(\cdot ,\zeta_0),\Psi_-(\cdot ,\zeta_0)).
    \end{equation}
  \end{itemize}
\end{Th}
\noindent We will use a second result concerning consistent solutions
near $\infty$. We won't need a basis in this case: a single solution
(actually the Jost solution) will be sufficient. It is given by
\begin{Th}[\cite{Kl-Ma:05a}]
  \label{thr:5}
  Assume (H1)-(H4) are satisfied. Fix $\eta\in\{+,-\}$ and
  $\alpha\in(0,1)$. Pick $E_0\in\R$ such that
  $E_0-W_\eta\not\in\sigma(H_0)$. Fix $X>1$ and $C_1>C_0$ (where $C_0$
  is defined in assumption (H4), section~\ref{sec:assumptions}).
  Then, there exist $C>0$ and $\delta>0$ such that, for any $E\in
  D(E_0,\delta)$, the cone $\mathcal{C}_\eta=\{\zeta\in\C;\
  \eta\text{Re}\,\zeta>C,\ C_1
  |\text{Im}\,\zeta|<\eta\text{Re}\,\zeta\}$ is a canonical domain.
  More precisely, fix $\zeta_0\in\mathcal{C}_\eta\cap\R$. Then, there
  exist a branch of the quasi-momentum
  $\kappa(\zeta)=\kappa(\zeta,E)$ that satisfies, for some
  $\sigma\in\{0,1\}$,
  \begin{equation}
    \label{eq:14}
    \forall(\zeta,E)\in\mathcal{C}_\eta\times D(E_0,\delta),\quad
    0<\text{Im}\,\kappa(\zeta,E)\quad\text{and}\quad
    \kappa^*(\zeta,E)=2\pi\sigma-\kappa(\zeta,E).
  \end{equation}
  and a function $(x,\zeta,E)\mapsto f(x,\zeta,E)$ defined on
  $(-X,X)\times\mathcal{C}_\eta\times D(E_0,\delta)$ that satisfies:
  \begin{itemize}
  \item $f$ is a consistent solution to equation~\eqref{G.2a}
  \item for $x\in(-X,X)$, $(\zeta,E)\mapsto f(x,\zeta,E)$ is
    analytic on $\mathcal{C}_\eta\times D(E_0,\delta)$;
  \item the function $f$ admits the asymptotics
    \begin{equation}
      \label{eq:15}
       f(x,\zeta,E)=e^{\eta\frac{i}{\varepsilon}
        \int_{\zeta_0}^{\zeta}\kappa(u,E)du}
      \left(\Psi_\eta(x,\zeta,E)+g(x,\zeta,E,\varepsilon)\right)
    \end{equation}
    where $(\Psi_+,\Psi_-)$ are the canonical Bloch solutions
    associated to $\kappa$ (see~\eqref{canonicalBS}) and $g$
    satisfy~\eqref{eq:47};
  \item the asymptotics can be differentiated once in $x$;
  \item the functions satisfy
    \begin{equation}
      \label{eq:13}
      \forall(x,\zeta,E)\in(-X,X)\times\mathcal{C}_\eta\times D(E_0,\delta),\quad
       f^*(x,\zeta,E)=e^{-2\pi\sigma\eta
         i(\zeta-\zeta_0)/\varepsilon}f(x,\zeta,E).
    \end{equation}
  \end{itemize}
\end{Th}
\noindent In~\cite{Mar:04,Mar:02}, a result analogous to
Theorem~\ref{thr:5} is proved, the main difference being the region in
$\zeta$ where the statements hold.
\subsection{The continuation diagrams}
\label{sec:cont-diagr}
Now to achieve the goal described in
section~\ref{sec:complex-wkb-method}, namely, to construct consistent
solutions to~\eqref{G.2a} with known asymptotics in large complex
domains of $\zeta$, as we have the form of standard asymptotics
locally both at a finite point in $\zeta$ and near infinity, we only
need patch together the various domain on which we found consistent
basis with standard asymptotics. An alternative way to obtain global
asymptotics was developed
in~\cite{MR2097997,MR2156718,MR2003f:82043,MR2091984}. It consists in
the proof of quite general continuation lemmas that describe geometric
situations in the complex plane of $\zeta$ under which one can prove
that, starting from a canonical domain and solutions with standard
asymptotics in this domain, one can continue them. The main obstacles
to continuation (i.e. to the validity of standard asymptotics) are the
nodal lines of $\im\kappa$ and the branching points of $\kappa$. We
will not discuss this further and refer to~\cite{Kl-Ma:05a} for the
details.

%
\def\cprime{$'$} \def\cydot{\leavevmode\raise.4ex\hbox{.}}

%
\end{document}